\documentclass[final,1p,times]{elsarticle} % do not change
\usepackage{graphicx} % do not change
\usepackage{amssymb} % do not change
\usepackage{amsthm} % do not change
\usepackage{lineno} % do not change

\usepackage[dvips]{hyperref}

\newcommand{\jpsi}{$J/\psi$}
\newcommand{\ee}{$e^+ e^-$}

\journal{Nuclear Physics A} % do not change
\begin{document} % do not change

\begin{frontmatter} % do not change

%% QM09Author: please enter your  
%% Title, author and address info here; please do not use footnotes

% Your Title - please modify
\title{\jpsi~and \ee~photoproduction measurements with PHENIX}

% Principle author, and co-authors - please modify
\author{Zaida Conesa del Valle$^{a}$ for the PHENIX Collaboration}

% Address - please modify
% note that if you have authors from several institutions, we recommend
% labelling these [a], [b], [c] etc.
\address[a]{
Laboratoire Leprince-Ringuet, Ecole Polytechnique - CNRS/IN2P3, Palaiseau, France
}

\hyphenation{a-gree-ment co-her-ent in-co-her-ent prom-is-ing}

\begin{abstract} % do not change
\jpsi~and \ee~photoproduction measurements in ultra-peripheral Au+Au collisions at $\sqrt{s_{NN}}=200$~GeV with the PHENIX experiment are presented~\cite{ppg081}. 
The triggered events correspond to those accompanied with forward neutron emission from the $Au^*$ dissociation. 
\jpsi~and high-mass \ee~cross sections and $p_T$ distributions are consistent with various theoretical calculations. 
\end{abstract} % do not change

\end{frontmatter} % do not change

%% QM09: we keep linenumbers at least for initial version
%\linenumbers % do not change

\section{Introduction} %: context and motivation}

\noindent
The photon is a fundamental ingredient of our understanding of elementary particles and their interactions~\cite{ppg081}. 
It can interact either directly (bare photon) or in a resolved manner, as a point-like particle (Compton-like scattering) or via quantum fluctuations to: a lepton pair, a quark anti-quark pair, or a vector boson with the same quantum numbers as the photon ($J^{PC}=1^{- -}$).
Photon-induced interactions have been traditionally studied with lepton beams (e.g. at PEP, LEP) and $\gamma p$ collisions (at HERA \& TeVatron). 
Ultra-peripheral collisions (UPC) of hadrons, proton or nuclei collisions where the impact parameter is larger than twice the nuclear radius such as there is no strong interaction, also allow the study of photon-induced reactions. 
The main advantatges of hadron vs. lepton colliders are the larger photon luminosity (the photon spectrum is proportional to $Z^2$, the two-photon luminosity goes as $Z^4$) and the opportunity to probe strong electromagnetic fields (coupling $\propto Z\sqrt{\alpha}$ instead of $\propto \sqrt{\alpha}$). 

\noindent
The $\sqrt{s_{NN}}=200$~GeV Au+Au UPC at the BNL RHIC collider provide means to test photon-nucleon and two-photon interactions at a maximum center of mass energy of $W_{\gamma n}^{max}\sim 34$~GeV and $W_{\gamma \gamma}^{max}\sim 6$~GeV respectively~\cite{ppg081}. This proceeding concentrates on the first measurement ever of $J/\psi \rightarrow e^+ e^-$~and high-mass \ee~in UPC heavy-ion collisions with the PHENIX experiment~\cite{ppg081,phenix}.
Exclusive \jpsi~photoproduction can proceed via Pomeron-exchange (two-gluon picture) either through coherent photon-nuclear ($\gamma A \rightarrow J/\psi$) or incoherent photon-nucleon ($\gamma n \rightarrow J/\psi$) reactions. The \jpsi~photoproduction cross section is then related to the gluon nuclear distribution functions, $G_A(x,Q^2)$. This allows one to constrain them in the small Bjorken-$x$ region, $x = m^2_{J/\psi}/W_{\gamma A}^2 \cdot e^{\pm y}$ which for $|y| \leq 0.35$ gives $x \in [7\cdot10^{-3}, 3\cdot10^{-2}]$, a relatively unexplored region (Fig.~\ref{fig:xvalues_phenix}~left) \cite{david_photon}, and to eventually probe quarkonia propagation in normal nuclear matter (the so called nuclear absorption). 
On the other hand, exclusive dilepton (\ee) pair production can occur through a pure electromagnetic process ($\gamma \gamma \rightarrow e^+ e^-$) and tests QED on a strongly interacting regime.

\section{Data analysis, results and interpretation}

\begin{figure}[!htbp]
\centering
\includegraphics[width=0.37\textwidth]{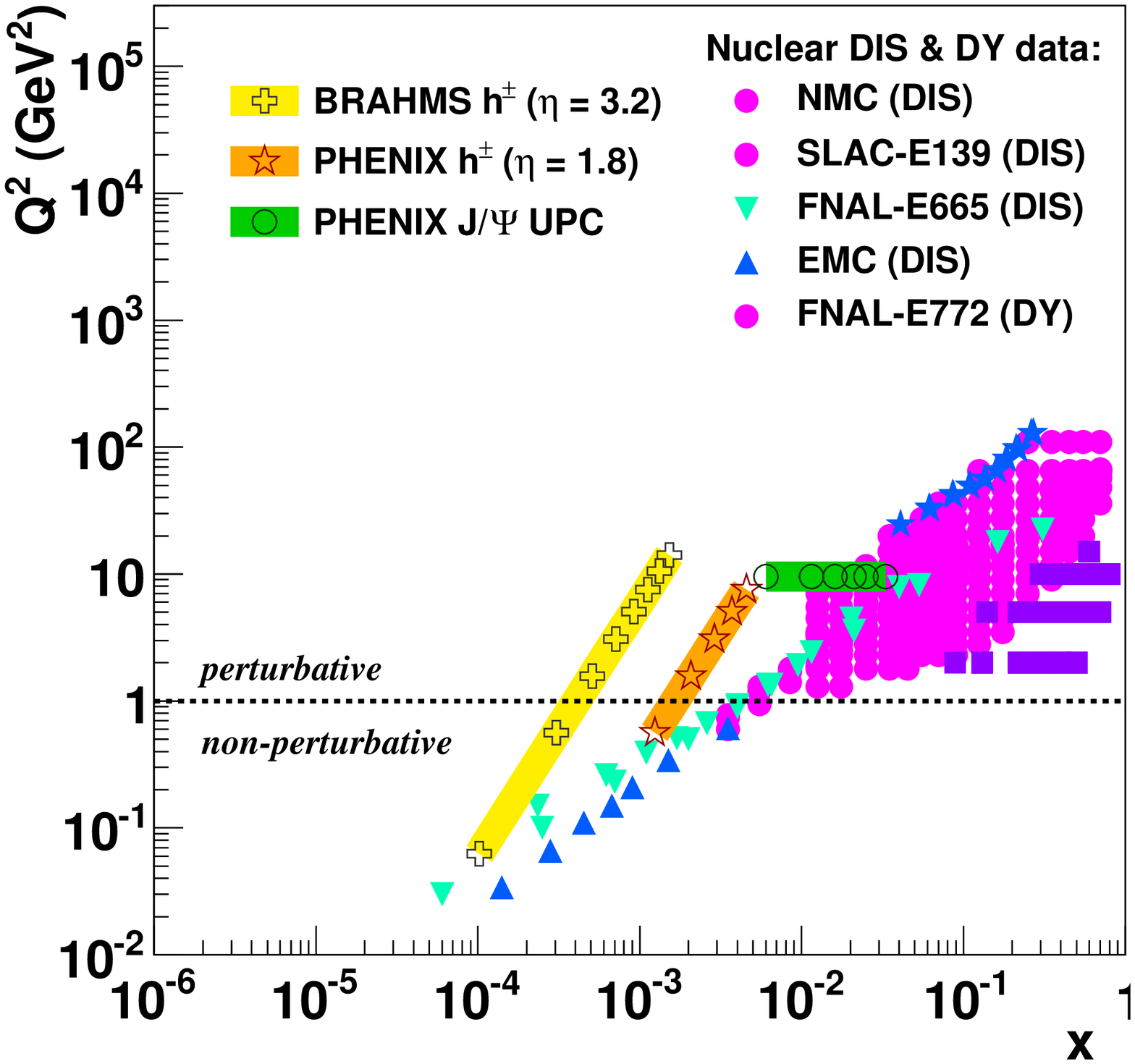}
\includegraphics[width=0.515\textwidth]{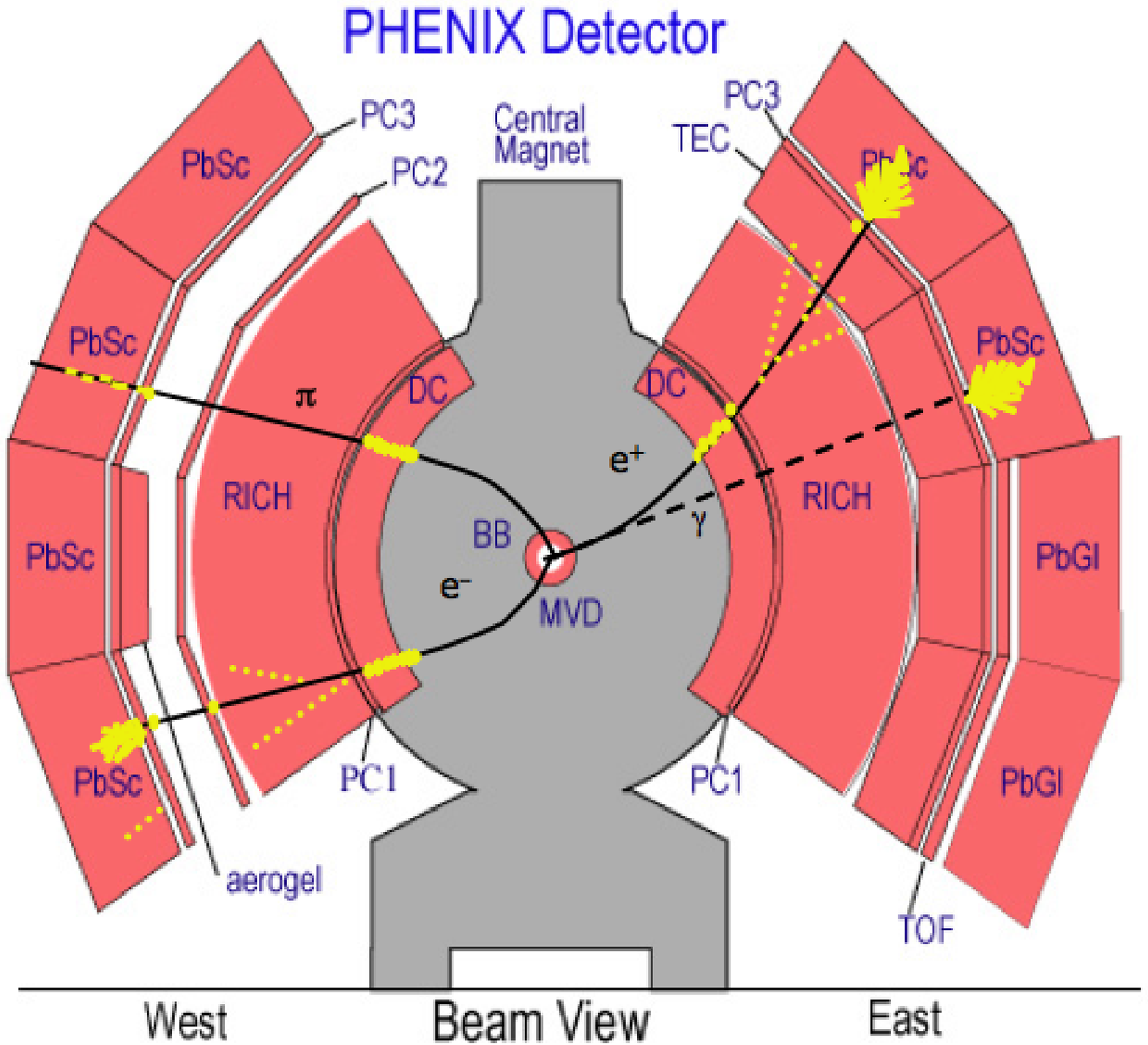}
\caption{
\label{fig:xvalues_phenix}
Left: Kinematic $Q^2$ versus Bjorken-$x$ map of the regions explored to probe the nuclear PDFs~\cite{david_photon}. 
Right: Section of the PHENIX central arms as they were on year 2004~\cite{phenix}. For illustration, passage of electrons, positrons, photons and pions through the detectors are depicted. 
}
\end{figure}

\noindent
% setup
The results summarized here~\cite{ppg081} concern data collected in 2004 on Au+Au collisions at $\sqrt{s_{NN}}=200$~GeV with the PHENIX experiment~\cite{phenix}. The apparatus is formed by four spectrometers (north, south, east, west) and a few global detectors. The central arms (east \& west spectrometers, Fig.~\ref{fig:xvalues_phenix}~right), are centered at mid-rapidity and measure electrons, photons and hadrons. The forward arms (north \& south spectrometers)  are for muon detection and identification. 
Electron tracking and identification is possible thanks to (from inner to outer): the multi-layer drift chambers (DC), the multi-wire proportional chambers (PC), the Ring-Imaging-Cherenkov detectors (RICH) and the electromagnetic calorimeters (EMCal) with two different technologies: lead-scintillador sandwich (PbSc), and lead-glass Cherenkov (PbGl) calorimeters.  

\noindent
Triggering on UPC interactions with such a detector, conceived to study the most central heavy-ion collisions, has been a successfully met challenge. 
The electromagnetic fields associated with these collisions can excite the nuclei, which will most probably de-excite by emitting neutrons at forward-rapidities.
The probability of \jpsi~coherent ($\gamma A$) or incoherent ($\gamma n$) photoproduction in coincidence with Au Coulomb neutron breakup amounts to $55\pm6\%$ or $\approx 100\%$ respectively~\cite{starlight}. 
The trigger strategy was then to: 
\begin{enumerate}
\item detect these neutrons in one or both of the Zero Degree Calorimeters (ZDC); 
\item reject non-UPC (peripheral or beam-gas) reactions by imposing a veto on coincident signals in both Beam Beam Counters (BBC), which also selects events with a large rapidity gap (BBCs coverage is of $3.0<|\eta|<3.9$); and
\item using the EMCal trigger to select events with one or more electrons of energy $>1$~GeV. 
\end{enumerate}

%  analysis
\noindent
The triggered events ($8.5$M events) were examined after applying data quality criteria ($21\%$ loss). The subsample of events analyzed are those centered in the detector fiducial area ($|z-vertex|\leq 30$~cm) and with only two charged tracks (electrons) to optimize the selection of exclusive processes. Electron identification criteria are also applied (see \cite{ppg081} for details). 

\begin{figure}[!htbp]
\centering
\includegraphics[width=0.495\textwidth]{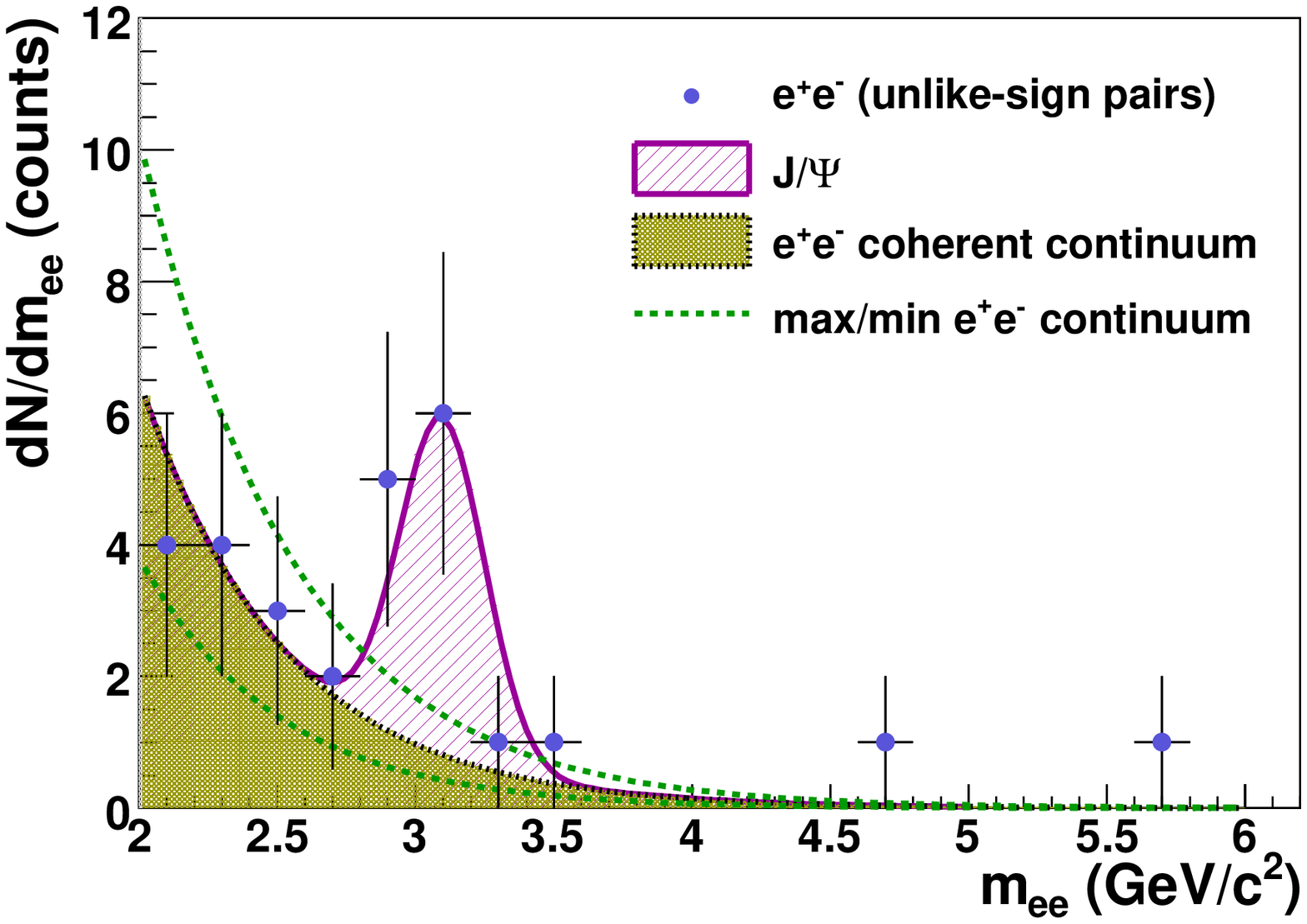}
\includegraphics[width=0.495\textwidth]{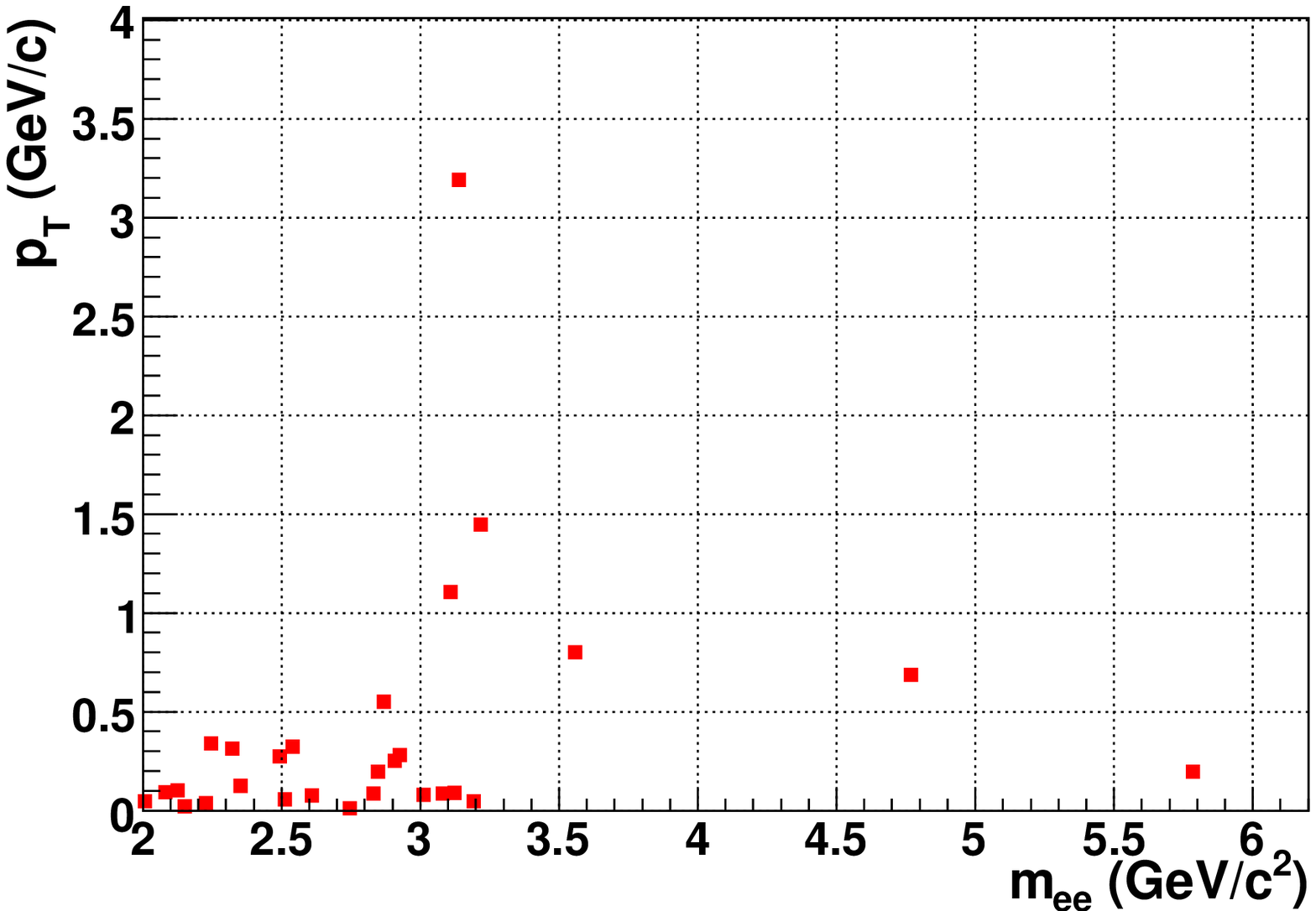}
\caption{
\label{fig:invmass_pt}
\ee~invariant mass distribution as measured in Au+Au UPC at $\sqrt{s_{NN}}=200$~GeV with PHENIX. 
The left plot depicts its decomposition into continuum and \jpsi. The right figure presents the invariant mass versus pair $p_T$ correlation. 
}
\end{figure}

% background
\noindent
A total of $28$ \ee~pairs and zero like-sign pairs ($e^+ e^+ + e^- e^-$) with $m_{ee}>2$~GeV/c$^2$ have been measured (Fig.~\ref{fig:invmass_pt} left). 
Despite the poor statistics, the trend is compatible with {\sc starlight}~\cite{starlight} MonteCarlo simulations and a full reconstruction with our experimental setup. 
The invariant mass distribution is fitted with a continuum (exponential) plus a \jpsi~(Gaussian) shape which allows one to deduce the number of \jpsi, $N[J/\psi]=9.9 \pm 4.1 {\rm (stat)}\pm 1.0 {\rm (syst)}$, and of continuum \ee~counts, $N[e^+ e^-]=13.7 \pm 3.7 {\rm (stat)}\pm 1.0 {\rm (syst)}$ for $m_{ee}\in[2.0,2.8]$~GeV/c$^2$. 
The statistical and systematical \ee~continuum uncertainties are computed by varying its subtraction method, changing its shape to a power law form and studying the dependence on the slope parameters. Those uncertainties are represented by dashed-lines in Fig.~\ref{fig:invmass_pt} left.

\noindent
The dielectron ($e^+ e^- + Xn$) cross section per invariant mass range is presented in Tab.~\ref{tab:xs_results}. 
Both, the pairs low transverse momentum, $p_T$, (Fig.~\ref{fig:invmass_pt} right) and 
the photoproduction cross sections, are in accordance with the {\sc starlight} predictions (Tab.~\ref{tab:xs_results}) for the coherent process, $\gamma \gamma \rightarrow e^+ e^-$. 
Consequently, \ee~photoproduction measurements agree with {\sc LO} QED theoretical ({\sc starlight}) calculations even in this strongly interacting regime. 
The caveats to this statement are the lack of comparisons with other predictions in the kinematical region of interest, and the fact that the {\sc NLO} corrections influence on the calculations remain uncertain~\cite{baltz}.

\begin{table}[!htbp]
  \begin{center}
    \begin{tabular}{ccc}
      \noalign{\smallskip} \hline\hline \noalign{\smallskip}
      $m_{e^+e^-}$ [GeV/c$^2$] & \multicolumn{2}{c}{$d^2\sigma/dm_{e^+e^-} dy|_{y=0}$ [$\mu$b/(GeV/c$^2$)]}  
      \\  
                  & {\sc data}      & {\sc starlight}
      \\  \hline
      $[2.0,2.8]$ & $  86 \pm  23\, {\rm(stat)} \pm 16 \, {\rm(syst)} $ & $90$ \\
      $[2.0,2.3]$ & $ 129 \pm  47 \, {\rm(stat)} \pm  28 \, {\rm(syst)}$ & $138$ \\  
      $[2.3,2.8]$ & $  60 \pm  24 \, {\rm(stat)} \pm  14 \, {\rm(syst)} $ & $61$ \\ 
      \hline\hline 
    \end{tabular}
  \end{center}
%\vspace{-6pt}
  \caption{
    \label{tab:xs_results}
    Measured $e^+ e^-$ continuum photoproduction cross sections at mid-rapidity in UPC Au+Au collisions (accompanied with forward neutron emission) at $\sqrt{s_{NN}}=200$~GeV. For comparison, the {\sc starlight} predictions are also quoted~\protect\cite{starlight}.
  }
\end{table}

\begin{figure}[!htbp]
\centering
\includegraphics[width=0.425\textwidth]{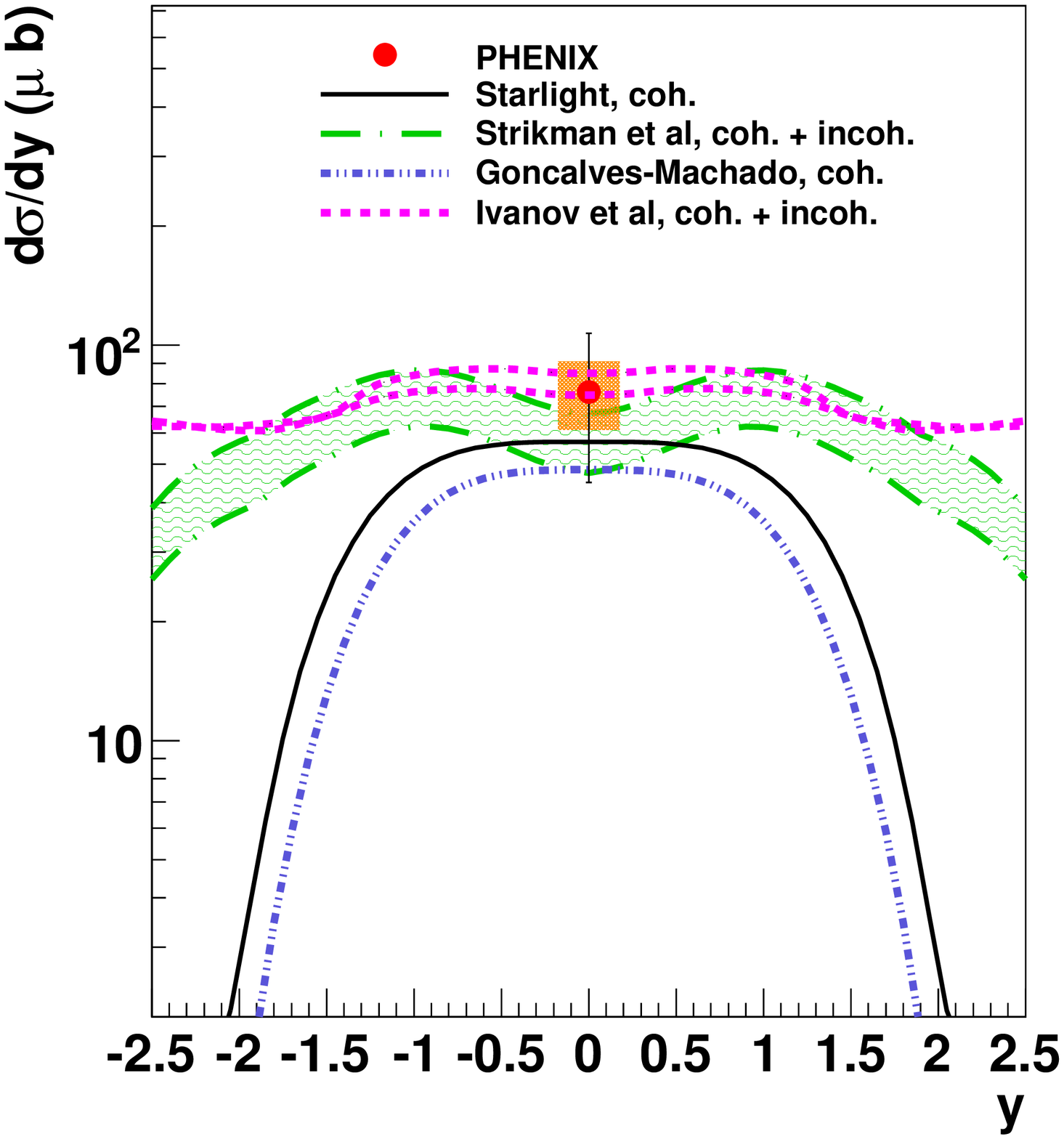}
\includegraphics[width=0.425\textwidth]{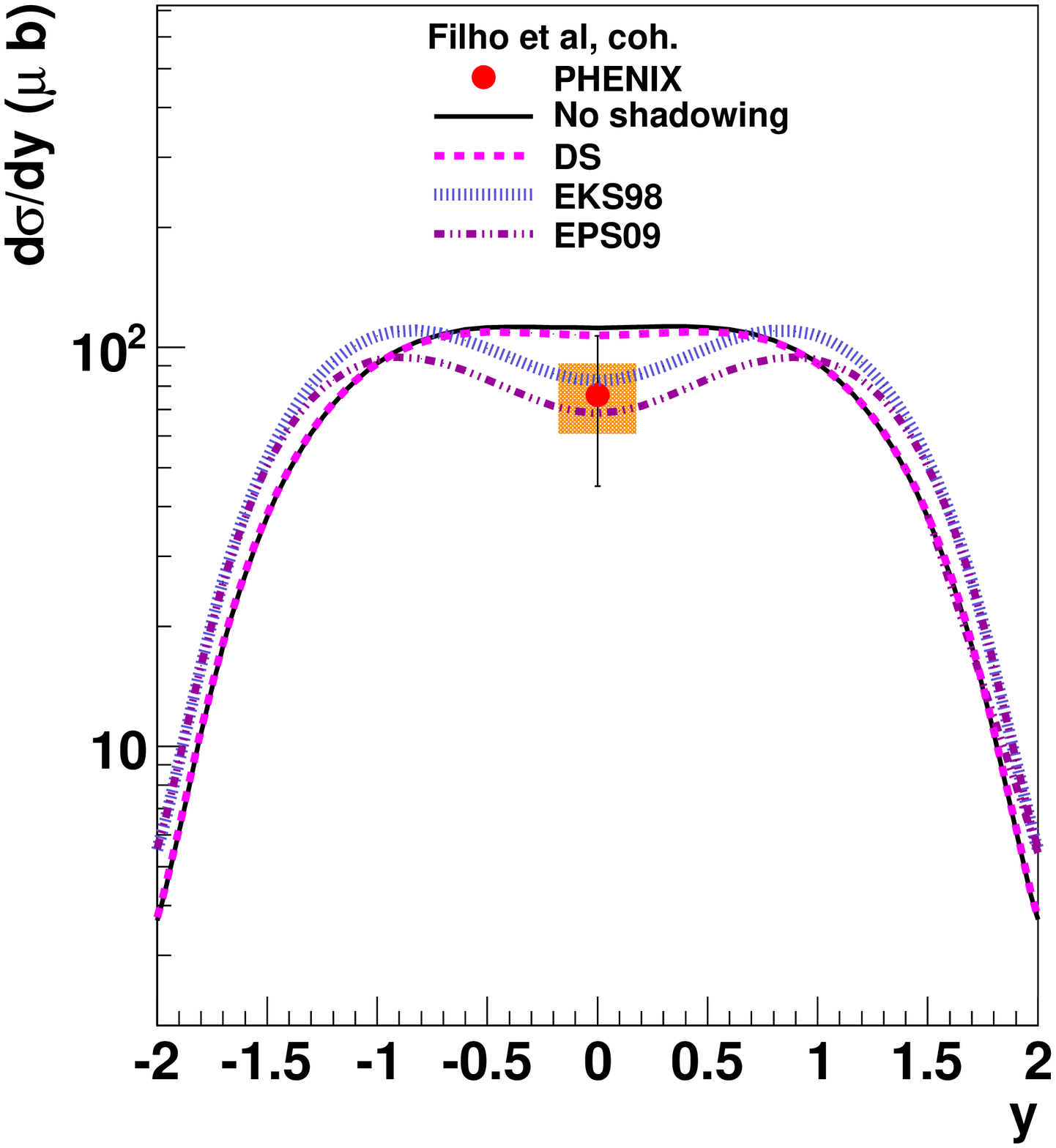}
% goncalves_shad_coh_prediction.eps}
\caption{
\label{fig:predictions}
Comparison of the measured $J/\psi+Xn$~photoproduction cross section with various theoretical calculations. 
Some of them include both coherent and incoherent components (labeled as ``coh. + incoh.'') : 
Strikman {\it et al}~\cite{strikman_pred}, Ivanov {\it et al}~\cite{ivanov_pred}.
The rest only compute the coherent contribution: {\sc starlight}~\cite{starlight}, Gon\c{c}alves {\it et al}~\cite{goncalves_pred}, Filho {\it et al}~\cite{filho_shadpred}.
}
\end{figure}

\noindent
The \jpsi~$p_T$ distribution (Fig.~\ref{fig:invmass_pt} right) suggests the presence of both coherent ($\gamma A$) and incoherent ($\gamma n$) contributions, consistent with low and intermediate to high $p_T$ respectively, and in agreement with the theoretical expectations~\cite{strikman_pred}. 
At mid-rapidity, the \jpsi~photoproduction cross section section accompanied with forward neutron emission ($J/\psi +Xn$) is 
 $d\sigma/ dy|_{y=0} \, [J/\psi] = 76 \pm 31 \, {\rm(stat)} \pm 15 \, {\rm(syst)} \, \, \mu$b.
This value is shown to be consistent with various theoretical calculations~\cite{starlight,strikman_pred,goncalves_pred,ivanov_pred,filho_shadpred} in Fig.~\ref{fig:predictions}.
Athough the current precision prevents more quantitative statements, the influence of the coherent and incoherent contributions to the predictions can be observed by comparing the results of calculations where both of them are included, labeled as ``coh.+incoh.'', and those where only the coherent component is considered, labeled only as ``coh.''.
In addition, Fig.~\ref{fig:predictions}~right illustrates the measurement's sensitivity to the nuclear PDFs by examining the Filho {\it et al} prediction~\cite{filho_shadpred} for coherent \jpsi~photoproduction with different shadowing parameterizations. 
Despite the poor statistics, which does not allow one to rule out any hypothesis, this is the first measurement of \jpsi~photoproduction in heavy-ion collisions and it shows itself to be a promising tool to learn about \jpsi~production processes and to probe the gluon nuclear PDFs at low Bjorken-$x$.

\section{Summary}

\noindent
The PHENIX experiment has proven its versatility by presenting the first measurement ever of high-mass \ee~and \jpsi~photoproduction in ultra-peripheral heavy-ion reactions~\cite{ppg081} in Au+Au collisions at $\sqrt{s_{NN}}=200$~GeV (accompanied by Au Coulomb nuclear breakup). 
Dielectron photoproduction cross sections are in agreement with theoretical {\sc starlight} {\sc LO} QED calculations for coherent two-photon production, $\gamma \gamma \rightarrow e^+ e^-$. 
\jpsi~photoproduction cross section and its $p_T$ pattern are consistent with the expectations~\cite{starlight,strikman_pred,goncalves_pred,ivanov_pred,filho_shadpred} and favour the possibility of both coherent ($\gamma A \rightarrow J/\psi$) and incoherent ($\gamma n \rightarrow J/\psi$) contributions. 
Even though the poor statistics prevents one from ruling out any hypothesis, it shows itself to be a promising tool to learn about \jpsi~production processes and probe the gluon nuclear PDFs at low-$x$. 
Future analysis at RHIC with higher luminosities and the imminent LHC collisions will provide more discriminating tools. 

%% end of main text

% \section*{Acknowledgments} % please check/modify, comment out or delete if not needed
% This is where one places acknowledgments for funding bodies etc., if needed.
% For the large collaborations, this is listed once and for all, together with 
% the author lists etc. in the proceedings back-material.

 % do not change 
\end{document}